%Paper: hep-th/9412056
%From: Tanya Khovanova <tanyakh@math.mit.edu>
%Date: Tue, 6 Dec 94 17:25:00 EST

%%%%%%%%%%%%%%%%%%%%%%%%%%%%%%%%%%%%%%%%%%%%%%%%%%%%%%%%
%                                                      %
%           On quantum group SL_q(2)                   %
%                                                      %
%       Joseph Bernstein, Tanya Khovanova              %
%                                                      %
%           plain TeX,   19 pages                      %
%                                                      %
%%%%%%%%%%%%%%%%%%%%%%%%%%%%%%%%%%%%%%%%%%%%%%%%%%%%%%%%

\def\date{le\ {\the\day}\ \ifcase\month\or janvier\or
{f\'evrier}\or mars\or avril \or mai\or juin\or juillet\or
{ao\^ut}\or septembre\or octobre\or novembre\or {d\'ecembre}\fi
\ {\oldstyle\the\year}}

\let\noi=\noindent

\let\what=\widehat

\def\a{\alpha} 
\def\g{\gamma}

\def\ve{\varepsilon}

\def\lda{\lambda}

\def\D{\Delta}

\def\L{\Lambda}
\def\Si{\Sigma}

\font\tenbb=msym10

\font\sevenbb=msym7
\font\fivebb=msym5

\newfam\bbfam
\textfont\bbfam=\tenbb \scriptfont\bbfam=\sevenbb
\scriptscriptfont\bbfam=\fivebb
\def\bb{\fam\bbfam}

\def\Z{{\bb Z}}
\def\C{{\bb C}}

\font\titre=cmbx12

\def\part{\partial}

\def \wtilde {\widetilde}

\def\ra{\rightarrow}
\def\longra{\longrightarrow}

\def\and{\mathop{\rm and}\nolimits}

\catcode`\@=11
\def\Eqalign#1{\null\,\vcenter{\openup\jot\m@th\ialign{
\strut\hfil$\displaystyle{##}$&$\displaystyle{{}##}$\hfil
&&\quad\strut\hfil$\displaystyle{##}$&$\displaystyle{{}##}$
\hfil\crcr#1\crcr}}\,} \catcode`\@=12

\catcode`\@=11
\def\displaylinesno #1{\displ@y\halign{
\hbox to\displaywidth{$\@lign\hfil\displaystyle##\hfil$}&
\llap{$##$}\crcr#1\crcr}}

\def\ldisplaylinesno #1{\displ@y\halign{
\hbox to\displaywidth{$\@lign\hfil\displaystyle##\hfil$}&
\kern-\displaywidth\rlap{$##$}
\tabskip\displaywidth\crcr#1\crcr}}
\catcode`\@=12

\def\buildrel#1\over#2{\mathrel{
\mathop{\kern 0pt#2}\limits^{#1}}}

\def\build#1_#2^#3{\mathrel{
\mathop{\kern 0pt#1}\limits_{#2}^{#3}}}

\def\hfl#1#2{\smash{\mathop{\hbox to 6mm{\rightarrowfill}}
\limits^{\scriptstyle#1}_{\scriptstyle#2}}}

\def\up#1{\raise 1ex\hbox{\sevenrm#1}}

\def\signed#1 (#2){{\unskip\nobreak\hfil\penalty 50
\hskip 2em\null\nobreak\hfil\sl#1\/ \rm(#2)
\parfillskip=0pt\finalhyphendemerits=0\par}}

\def\TeX{T\kern-.1667em\lower.5ex\hbox{E}\kern-.125em X}

\def\lsim{ {\raise -3mm \hbox{$<$} \atop \raise 2mm
\hbox{$\sim$}} }

\def\gsim{ {\raise -3mm \hbox{$>$} \atop \raise 2mm
\hbox{$\sim$}} }

\def\frac#1#2{\mathop{\scriptstyle#1\over\scriptstyle#2}\nolimits}

\def\fnote#1{\advance\noteno by 1\footnote{$^{\the\noteno}$}
{\eightpoint #1}}

\def\boxit#1#2{\setbox1=\hbox{\kern#1{#2}\kern#1}%
\dimen1=\ht1 \advance\dimen1 by #1 \dimen2=\dp1 \advance\dimen2 by
#1
\setbox1=\hbox{\vrule height\dimen1 depth\dimen2\box1\vrule}%
\setbox1=\vbox{\hrule\box1\hrule}%
\advance\dimen1 by .4pt \ht1=\dimen1
\advance\dimen2 by .4pt \dp1=\dimen2 \box1\relax}

\def\cube{
\raise 1 mm \hbox { $\boxit{3pt}{}$}
}

\def\cqfd{\unskip\kern 6pt\penalty 500
\raise -2pt\hbox{\vrule\vbox to10pt{\hrule width 4pt
\vfill\hrule}\vrule}\par}

\def\dstar {\displaystyle ({\raise- 2mm \hbox
{$*$} \atop \raise 2mm \hbox {$*$}})}

\def\ref #1#2{
\smallskip\parindent=1,0cm
\item{\hbox to\parindent{\enskip\lbrack{#1}\rbrack\hfill}}{#2} }

\def\choose#1#2{\mathop{\scriptstyle#1\choose\scriptstyle#2}\nolimits}

\def\adots{\mathinner{\mkern2mu\raise1pt\hbox{.}
\mkern3mu\raise4pt\hbox{.}\mkern1mu\raise7pt\hbox{.}}}

\def\pegal{\mathrel{\vbox{\hsize=9pt\hrule\kern1pt
\centerline {$\circ$}\kern.6pt\hrule}}}
\magnification=1200
\overfullrule=0mm

\def\Mat{\mathop{\rm Mat}\nolimits}

\centerline {\titre On quantum group $\bf SL_q(2)$.}
\vglue 1cm
$$\displaylines{
{\rm Joseph\  Bernstein} \cr
\hbox{\rm Tel-Aviv  University} \cr
\hbox{\rm Tanya  Khovanova} \cr
\hbox{\rm MIT } \cr
}$$

\vglue 1.5cm
\noi {\bf Introduction.}
\medskip
The goal  of  this  paper  is to analyze the notion of quantum
group. There are two  approaches  to  this  notion:
\medskip
In  first
approach, one  describes a
quantum group  $G$  in terms of a Hopf algebra $A = A(G)$ which
plays the role of the algebra of functions on $G$. Then one
studies   the   monoidal   category   of
$A$-comodules, which  is thought of as the category of representations of $G$.
Our basic motivating example is the  algebra  $A$  of  regular
functions on quantum group $SL_q(2)$
(see [R-T-F], [M]).
\medskip
In second approach, one describes the quantum group in terms of
a  Hopf  algebra  $U  \subset
A^*$, which  plays the role of universal enveloping algebra,
and studies the tensor category of  $U$-modules.  This  approach
was initiated   by   Drinfeld [D] and  Jimbo [J].  We  use  Lusztig's
exposition (see [L]).
\bigskip
We decided to find a way from  the  first  definition  to  the
second one trying  to  understand  what  axiomatic and structures
lied behind it.
\medskip
We begin  with  the  Hopf  algebra  $A = A_q$
of  regular  functions  on
$SL_q(2)$. This Hopf algebra supplies us with the material for
axiomatic construction and generalizations. Then our axiomatic
approach gives  us  the  direction  in  which  to make further
investigation of the Hopf algebra of regular functions on $SL_
q(2)$, and so on.
\bigskip
The results  of this article could be formulated as follows:
\medskip
1. We propose an axiomatic construction of Hopf  algebras  $A$
and $U$,  based  on the observation that the quantum group $G$
contains a quantum subgroup,  which is  just  a  usual  Cartan
subgroup. In  this  paper we give detailed analysis of $SL_q(2)$
case,  starting from the one-dimensional torus
and the root system of $SL(2)$.
\medskip
2. Our construction leads us to a new class of quantum groups
of $SL(2)$-type  which  seem  to  be  the quantum analogues of
metaplectic extensions of the group $SL(2)$.
\medskip
3. Our  approach  forced  us  to  introduce a notion   of
tetramodule (see [Kh]). We are thankful to V.Lyubashenko and S.Shnider who
pointed out to us that this notion had already appeared in literature
under many other names. For example: bidimodule [G-S], Hopf bimodule
[L-S], two-sided two-cosided Hopf module [Sch], 4-module [S-S],
 bicovariant bimodule [W].
\bigskip
We plan to develop similar approach to other  quantum  groups
in future publications.
\bigskip
     This work  was partly done  during  our   visit   to   Tel-Aviv
University and I.H.E.S.. The authors thank Department of Mathematics of Tel
-Aviv University and I.H.E.S.  for supporting this  work.

The first
author was partially supported by an NSF grant.
The second author was supported by the NSF Grant No. DMS - 9306018.
\medskip
The second author is thankful to V.Kac, D.Kazhdan,
M.Khovanov and L.Korogodsky for their interest
in this paper and stimulating discussions, to V.Lyubashenko
for his help with literature and references and useful discussions.
\bigskip
\noi {\bf 1. Algebra of regular functions on $\bf SL_q(2)$.}
\medskip
{\bf 1.1.} For every $q \in \C^*$ we consider the quantum group $SL_q(2)$.
This group is defined by its algebra of functions $A_q$ which is an algebra
generated by four noncommuting elements $(a,b,c,d)$, satisfying the
following
relations [R-T-F]:
$$\Eqalign {
ab & = q^{-1}ba \qquad & ac       & = q^{-1}ca      \cr
cd & = q^{-1}dc        & bd       & = q^{-1}db      \cr
bc & = cb              & ad - da  & = (q^{-1}-q)bc  \cr
 }\leqno (*)$$
$$ad - q^{-1}bc = 1$$
\medskip
Introduce matrices
$$ \Eqalign {
Y &= \pmatrix { a & b \cr c & d \cr } \in {\rm Mat}\ (2,A_q) \cr \cr
Y^t &= \pmatrix
{a & c \cr b & d \cr } \qquad Q = \pmatrix { 0 & -1 \cr q^{-1}
& 0 \cr }}. $$
Then we can rewrite the relations $(*)$ in a more compact form:
$$ \eqalign {
Y Q Y^t &= Q \cr
Y^t Q Y &= Q\ . \cr
}$$
\medskip
{\bf 1.2.} {\it  Remark.}
In a similar way one can describe the quantum group $GL_q(2)$.
Namely,
the relations
$$\matrix {
\eqalign {Y Q Y^t &= x_1 Q \cr
Y^t Q Y &= x_2Q \cr} &\quad x_1,x_2 \in \C^* \cr
}$$
imply that $x_1 = x_2$. Hence, we can define the algebra of functions
on the quantum group $GL_q(2)$ as the algebra generated by $a,b,c,d,x$
with relations: 1) $x$ is invertible element in the center;
2) $Y Q Y^t = x Q = Y^t Q Y$.
\medskip
{\bf 1.3.} The comultiplication in the algebra $A_q$ is defined as follows:
$$\eqalign {
\D a &= a \otimes a +  b \otimes c    \cr
\D b &= a \otimes b +  b \otimes d    \cr
\D c &= c \otimes a +  d \otimes c    \cr
\D d &= c \otimes b +  d \otimes d \ .\cr
} \leqno (**)$$
Using the natural imbeddings $i',\ i'' : A_q \ra A_q \otimes A_q,\ \bigl (
i' (x) = x \otimes 1,\ i''(x) = 1 \otimes x \bigr )$, we can rewrite
comultiplication formulae $(**)$ as follows:
$$\D(Y) = i'(Y) \cdot i''(Y)\ ,$$
which is an equality in $\Mat (2,A_q \otimes A_q)$.
\medskip
There exists an antipode $i$ in the algebra $A_q$, and it
is defined as follows:
$$\eqalign {
a \mapsto d & \ \ \ b \mapsto -q^{-1}b \cr
d \mapsto a & \ \ \ c \mapsto -qc\ . \cr
} $$
In a more compact form:
$$ Y \mapsto (Q^{-1}YQ)^t.$$
\medskip
{\bf 1.4.} Analyzing this algebra we note that it has the following
important property:
\medskip
Let $I \subset A_q$ be a two-sided ideal generated by $b $ and $c$.
Then $I$ is a
Hopf ideal in $A_q$, i.e. $\D I \in A \otimes I + I \otimes A$ and
$i(I) \subset I$.
The quotient Hopf algebra $S = A_q/I$ is
isomorphic to the algebra of functions on the algebraic group $H = \C^*$:
$$S = \C [a,d] \big  / (ad-1) $$
$$\D a = a \otimes a \ \ \ \D d = d \otimes d $$
\medskip
Informally, this means that our quantum group $SL_q(2)$ contains
the usual group $H = \C^*$ as a subgroup.
\bigskip
\noi {\bf 2. Axiomatic approach.}
\medskip
{\bf 2.1.} Let us fix  a  torus $H$ (i.e.  an  algebraic  group
isomorphic to  $\C  ^{*n}$)  and  denote by $S$ the Hopf algebra of
regular functions on $H$.  We would like to study pairs  $(A,I)$
satisfying the following property:
\medskip
    {\it Assumption I}.
 $A$ is a Hopf algebra (with multiplication $m$ and
comultiplication $\D$) and $I  \subset  A$ is a  two-sided  Hopf
ideal, such that $A/I$ is isomorphic to $S$ as a Hopf algebra.
\medskip
{\bf 2.2.} Note that the comultiplication $\D : A \ra A \otimes A$
leads
to two $S$-comodule structures on $A$:
$$\displaylines {c_\ell : A \ra S \otimes A \qquad c_\ell = (pr \otimes id) \D
\cr c_r : A \ra A \otimes S \qquad c_r = (id \otimes pr) \D \cr
}$$
where $pr$ is the natural projection $A \ra S = A/I$.
\medskip
These structures define algebraic actions $s_\ell$ and $s_r$ of the
algebraic
group $H$ on $A$. These actions commute and preserve $I$.
\medskip
We also consider the adjoint action of $H : ad_h = s_\ell(h) \cdot
s_r^{-1}(h)$, $h \in H$.
\medskip
{\bf 2.3.} Consider an associated graded algebra $gr\, A$:
$$gr\, A = \build{\oplus}_{0 \le n}^{} gr_n\ A, $$
where $$gr_n \,A = I^n/I^{n+1}\ .$$
It is easy to see that $gr\, A$ inherits the structure of a graded Hopf algebra
with $gr_0 A$ equal to $S$. In particular, $grA$ has the structure of
a graded $S$-bicomodule.
\medskip
{\bf 2.4.} We have two natural $S$-module structures on $grA$. These
structures commute and preserve $gr_n A$.
\medskip
The $S$-bicomodule and $S$-bimodule structures are compatible. We will
describe,
for example, the compatibility of the left $S$-comodule and right $S$-module
structures: for any $h \in H,s \in S,\ x \in gr A$
$$s_\ell(h)(xs) = s_\ell(h)(x) \cdot s_\ell(h)(s),$$
or, in other words,
$$c_\ell(xs) = c_\ell(x) \D s\ .$$
The other three relations are of the same type:
$$c_\ell(sx) = \D s\cdot c_\ell(x)$$
$$c_r(xs) = c_r(x) \D s$$
$$c_r(sx) = \D s\cdot c_r(x).$$
\medskip
{\bf 2.5.} {\it Definition}. We call the linear space $V$ an $S$-{\it
tetramodule} if $V$ is equipped with commuting left and right $S$-module
structures, commuting left and right $S$-comodule structures, and
the $S$-bimodule and $S$-bicomodule structures are compatible (see 2.4).
\bigskip
\noi {\bf 3. $\bf S$-tetramodules.}
\medskip

{\bf 3.1.} For properties of tetramodules
we use notations of [Kh]. See also [A], [S], bidimodules - [G-S],
Hopf bimodules - [L-S], two-sided two-cosided Hopf modules - [Sch],
4-modules - [S-S], bicovariant bimodules - [W].
\medskip
{\bf 3.2.} There is another shorter and more invariant way to
describe the notion of tetramodule.
   Let us define a (0,1)-graded space to be a graded vector space
  $ B = \oplus B_i$ such that $ B_i = 0$ for $i \neq 0, 1 $. In the
  category of (0,1)-graded spaces we can define the restricted
  tensor product by
  $U {\otimes}_r  V  =
  (U \otimes V)_0 \oplus (U \otimes V)_1 =
  U \otimes V / (U \otimes V)_2 $.
\medskip
  Now define a restricted bialgebra as a (0,1)-graded vector space $B$
  together with morphisms
  $m : B \otimes_r B \to B $, $ \Delta: B \to B \otimes_r B$,
  $\ve: \C \to B $ and $ \eta: B \to \C$,
  satisfying usual axioms of bialgebra.
  \medskip
   {\it Lemma}. Let $B = (B_0 = S, B_1 = V)$ be a (0,1)-graded vector
   space. Then to define a structure of restricted bialgebra on $B$ is
   the same as to define a structure of bialgebra on $S$ and a
   structure of $S$-tetramodule on $V$.
\medskip
  {\bf 3.3.}
Consider the case when  $S$  is  the  Hopf algebra  of
regular functions on a torus $H$.  In this case we can give an
explicit description  of the category of $S$-tetramodules (for any
$S$, see [A], [S]). We
use the following standard
\medskip
{\it Lemma.} Let $W$ be an $S$-module equipped with the compatible
algebraic action of the group $H$. Then $W = S \otimes W^H$, where $W^H$ is
the space of $H$-invariants.
\medskip
{\bf 3.4.}
Let us apply this lemma to our case. Let $V$ be an $S$-tetramodule.
Applying lemma 3.3 to the right action of $H$ on $V$ and the
right multiplication by $S$ we can write $V$ as
$V = V^H \otimes S$.
\medskip
  We want to describe an $S$-tetramodule structure on $V$
in terms of some structures on the vector space $V^H$.
\medskip
The right action of $H$ on $V^H$ is trivial. It is clear that $V^H$ is
$ad_H$-invariant, so the left action of $H$ on $V^H$ coincides with the $ad_H$
action. Hence, knowing the $ad_H$ action on $V^H$, we can reconstruct the left
and right actions of $H$ on $V$.
\medskip
The right action of $S$ on $V$ is defined by decomposition $V = V^H \otimes
S$. Now we have to reconstruct the left action of $S$ on $V$.
\medskip
Let $\L$ be the lattice of characters of $H$. Then $\L \subset S$ is a basis of
$S$. For $\lda \in \L$ consider operators $m_\ell(\lda)$ and
$m_r(\lda)$ of left and right multiplications on $\lda$ in $V$,
and set  $L(\lda)   =   m_\ell(\lda)   m_r(\lda)^{-1}$.   Then
the axiom H3 implies that the operators $L(\lda)$
commute with  the  right and the left action of $H$ and hence
preserve the subspace $V^H$.
\medskip
So we have defined a homomorphism $L$ of $\L$ into automorphisms of $V^H$,
commuting with $ad_H$. Knowing  $L$ we can reconstruct the left action of $S$
on $V$.
\medskip
{\bf 3.5.}  {\it  Summary.}  Let  $S$  be  the Hopf algebra of
regular functions on a torus $H$. Then the functor $V \to V^H$
gives an equivalence of
the category of $S$-tetramodules with the category of
algebraic $H$-modules equipped with the commuting action $L$ of the lattice
$\L$ .
\bigskip
\noi{\bf 4. Next step -- $\bf A/I^2$.}
\medskip
{\bf 4.1.} Let us return to a Hopf algebra  $A$  with  a  Hopf
ideal $I$, such that
 $ gr_0\, A$ equals $S$. We denote $gr_1 A$ by $T$. Then $T = I/I^2$ is an
$S$-tetramodule. Let us denote $T^H$ by $M$.

  We assume that our tetramodule T satisfies
the following assumption:
\medskip
  {\it Assumption II}.
    The space $M$ is a direct sum of nontrivial nonequivalent
one-di\-mensional representations of torus $H$.

\medskip

{\bf 4.2.} {\it Example.}
Let $G$ be a reductive algebraic group, $H$ its Cartan subgroup. Let $A =
\C[G]$ be the Hopf algebra of regular functions on $G$ and $I$ the
 ideal of
functions equal to $0$ on $H$. Then $S = A/I$ equals $\C [H]$,
$ T = I/I^2$
is an $S$-tetramodule. The space $M = T^H$ is
isomorphic to $({\cal G/H})^*$.
\medskip
As an $H$-module, $M$ is a direct sum of one-dimensional representations
$M_\alpha$  which
correspond to roots of $G$; in particular it satisfies
Assumption II.
\medskip
   Consider a family of quantum deformations $ G_q$ of the group
   $G$. By this we mean a flat family of Hopf algebras $A_q$
   depending on some parameter $q$ which for some value of $q$ gives
   $A$. Let us assume that we also can flatly deform the ideal $I$
   such that the family of quotient Hopf algebras $H_q = A_q / I_q $
   is constant and equals $S = \C [H]$.
     Under such deformations dimensions of different components of
     the space $M_q$ can only drop, so it will satisfy Assumption
     II.
     If we take, for example, the Hopf algebra of functions on
     $SL_q(n)$
(see, for example [M2]) for generic $q$ the space $M$ will
   have only components $M_\alpha$ corresponding to roots $\alpha$
   such that either $\alpha$ or $-\alpha$ is a simple root (see [Kh]).

\medskip
{\bf 4.3.}
Now we consider an $S$-tetramodule $T$
satisfying the assumption II.
Thus $M = \oplus M_\a$, where $\a$ runs some finite subset $\Si
\subset \L \setminus \{ 0 \}$ and $\dim M_\a = 1$ for every $\a$.
The action $L$ of the lattice
$\L$ on $M$ commutes with $ad_H$ and hence preserves
every subspace $M_\a$. On the space
$M_\a$ a homomorphism $L$ (see 3.4) is given by a
character $\g_\a : \L \ra \C^*$ that is by an
element $\g_\a$ in $H$.
\medskip
{\bf 4.4.} {\it Summary.}
Under assumptions I and II  an $S$-tetramodule T is fully described
up to an isomorphism by a finite subset $\Si$ in $\L
\backslash \{ 0 \}$ and a map $\g : \Si \ra H$.
\medskip
{\bf 4.5.}
Consider the adjoint action of the group $H$ on the
exact sequence
$$0 \ra T \ra A/I^2 \ra S \ra 0\ .$$
Then the action on $S$ is trivial and the action on $T$ by
assumption II
does not have invariant vectors. Thus as an $H$-module this sequence
canonically splits: $A/I^2 = S \oplus T$.
\medskip
{\bf 4.6.}
{\it Resume.}  Under  assumptions I and II   an  $S$-tetramodule $T$
allows us to describe
completely the   structure   of   $A/I^2$  as  an  algebra  and
an $S$-bicomodule.
\bigskip
\noi {\bf *. Passing to the dual picture.}
\bigskip
\noi {\bf *.1. General case.}
\medskip
{\bf *.1.1.} Let $A$ be a Hopf algebra. We would like to describe the
monoidal category  ${\rm Rep}(A)$ of representations of A, i.e.
monoidal category of left $A$-comodules.
    One of the standard ways to do it is to pass to modules over
the dual algebra $A^*$.
    We define the  multiplication in the vector space $A^*$ to be dual
 to the opposite comultiplication in $A$: for
 $f,g \in A^*$ we define $f \cdot g \ (a)$ as $(g \otimes f)
\ (\D(a))$.
\medskip
The antipode can also be easily defined: $i(f)(a) = f (i(a))$.
\medskip
     Let $\rho : V \ra A \otimes V$ be a left $A$-comodule. Then
every functional $f \in A^*$ defines
an operator $\rho(f) : V \ra V$ as a through
map $V \build {\longra}_{}^{\rho} A
\otimes V \build {\longra}_{}^{f \otimes id} V$.
It is clear that $\rho(f) \cdot \rho(g)$ equals $\rho (f
\cdot g)$, i.e. a left $A$-comodule defines a left $A^*$-module
(this is the reason why we prefer the multiplication in $A^*
$ to be {\it opposite} to the comultiplication in $A$).
So we have described a fully faithful functor from ${\rm Rep} (A)$
into the category of left $A^*$-modules.
\medskip
{\bf *.1.2.} If $A$ is finite dimensional we define the comultiplication in
$A^*$ to be dual to the opposite multiplication in $A$:
$$\D^*f (x \otimes y) = f(yx) \leqno(*).$$
Then $(A^*,  m^*,  \D^*)$ is  a  Hopf
algebra.
\medskip
When $A$ is infinite dimensional the formula $(*)$ defines $\D
^*f$ as an element in $(A \otimes A)^*$ which is bigger than $
A^* \otimes A^*$.  Thus we can not consider $A^*$  as  a  Hopf
algebra.

There are two usual strategies how to deal with this difficulty.
First is to
choose a  completion  of  $A^*  \otimes A^*$ such that $\D^*f$
would be defined for every $f \in A^*$.  Second is to choose  a
subalgebra $U \subset A^*$ on which $\D^*$ is defined,
which is closed with respect to $\D^*$ and which  is  "big
enough" (this means that $U$ is dense in $A^*$ in weak topology, or,
equivalently, that the orthogonal complement of $U$ in $A$ is
$0$).
\medskip
Our approach is to construct a Hopf subalgebra $U \subset A^*$
and describe some subcategory of $U$-modules which is close enough
to  the category ${\rm Rep} (A)$.
\bigskip
\noi {\bf *.2. Basic example.}
\medskip

{\bf *.2.1.} We consider the following simple but very instructive
case.
   Let $A = S = \C [H]$ be the Hopf algebra of regular functions on
  a torus  $H$. Then the dual algebra $S^*$  can be realized as the
  algebra of all
  functions on the lattice $\L$ of characters of $S$.
  There are several natural choices for a subalgebra $U \subset S^*$:

   (i) Algebra $U = U^0$ - a free algebra generated by elements $\hat h
  \  (h \in H \subset S^*)$.
This is a Hopf subalgebra, since $\Delta(\hat h) = \hat h \otimes \hat h$.
  Here we use $\hat h$ to differentiate the generator of an
algebra from the point of
  a torus. Our example - $SL(2)$ - could be the most confusing since
points of
  the torus are described by non zero complex numbers.
  \medskip
  In fact, they often use even smaller subalgebra, generated by some
  subgroup of $ H$. For example, in $SL(2)$ case we can take
  the smaller subalgebra
  generated by $\hat q, \hat q^{-1}$.

  (ii) $U = U(\cal H)$ - the enveloping algebra of the Lie algebra of
  $H$. This is a Hopf subalgebra.

  (iii) $U = U^f$ - subalgebra generated by $U^0$ and $U(\cal H)$. This
subalgebra can be described as the algebra of all functionals, which
are finite with respect to multiplicative action of $S$.
This is a Hopf subalgebra.

  (iv) $U = S^*$. In this case we have to complete $S^* \otimes
  S^*$
   to $(S \otimes S)^*$ in order to be able to define $\Delta$.
We set $S^* \hat \otimes S^* := (S \otimes S)^*$.
   The algebra $S^* \hat \otimes S^*$ could be realized
as the algebra of
   all functions on the lattice $\L \oplus \L$.
   If $f \in S^*$ then $\Delta f (\lambda_1, \lambda_2) = f (\lambda_1 +
   \lambda_2)$.
   \medskip
   {\bf *.2.2.} Our main interest is the category of $S$-comodules =
   the category of algebraic representations of $H$.
   We can describe the category of algebraic representations of $H$
   as the category of $U$-modules, which are algebraic when
   restricted to $H$ (when $H$ is not a subalgebra
   of $U$ we should say correspond to algebraic representation of $H$).
   \medskip
Thus, from our point of view, the dual object to $S$ is any Hopf algebra
   $U$ which is "dual" to $S$ in the sense described above {\it
   together} with a category of $U$-modules which are algebraic when
   restricted to $H$.
   \medskip
   {\bf *.2.3.} Later in this paper we prefer to take
   $U = S^*$ as this is the most general case; and
   it includes all other cases; or, alternatively one can take
 $U = U^0$ - in this case
   comultiplication formulae look clearer and simpler.
   \bigskip
   \noi {\bf *.3. Axiomatic approach.}
   \medskip
   {\bf *.3.1.}
Let $A$ be a Hopf algebra with a Hopf ideal $I$,
satisfying the assumption I.
\medskip
      We call a functional $f \in A^*$ $I$-finite
if it vanishes on some power of the ideal $I$. The space of $I$-finite
functionals we denote by $U$. This space is an algebra;
and has a natural algebra filtration
$U_0 \subset U_1 \subset \dots$, where $U_i$ consists of
functionals which vanish on $I^{i+1}$.
   In particular, the subalgebra  $U_0 = S^*$ contains $H$ as a
   subgroup.
\medskip
 {\it Definition.} The $( U, H)$-module is a $U$-module
such that its restriction to $H$ is an algebraic representation of $H$.
The category of $( U,H)$-modules we denote by ${\cal M} ( U,H)$.
   As follows from the construction we have a canonical faithful
   functor ${\rm Rep}(A) \to \ {\cal M}(U,H)$. This functor is fully
   faithful provided $U$ is dense in $A$, i.e. provided that the
   powers of the ideal $I$ have $0$ intersection.
\medskip
{\bf *.3.2.}   Let us describe in more detail the structure of the algebra $U$.
   We saw that $U_0 = S^*$. Also, it is clear that $U$ is generated by
   $U_1$ as an algebra.
      In order to describe the structure of $U_1$ we suppose
      that $A$ satisfies the assumption II. That means we can use
      decomposition $A/I^2 = S \oplus T$ from 4.5.
\medskip
     For every $\a \in \Si$, we fix a non-zero vector $E_\a$ in the
one-dimensional space $M^*_{-\a}$. Since $T = \bigl
( \build{\oplus}_{\a\in \Si}^{} M_\a \bigr ) \otimes S$, then
$E_\a$ defines a morphism $S^* \ra T^* \subset A^*$.
In particular, it defines a family of elements $E_\a(f) \in T^*$
for $f \in S^*$.
Now the $S$-bicomodule structure of $T$ defines
an $S^*$-module structure on the space of operators
spanned by $E_\a(f)$. It is easy to describe this structure:
$$\eqalign {
E_\a(f_1)f_2 & = f_2(\a)E_\a(f_1f_2) \cr
f_2E_\a(f_1) & = E_\a(f_1f_2) \ .
}$$

    It is clear, that if for every $\a$ we fix $f_\a \in S^*$, then
    $U_1 = S^* \oplus \oplus_\a S^* E_\a(f_\a)$.
    \medskip
    {\bf *.3.3.} Now we want to define the comultiplication on $U$.
In order to do this, we need to complete $U \otimes U$. We set
$$U \hat \otimes U := (S^* \hat \otimes S^*) \otimes_{S^* \otimes S^*}
 (U \otimes U).$$
       The comultiplication $\Delta$ on $S^*$ is defined as in *.2.1.(iv).
       We also have
       $$\D(E_\a(f)) = \D f \cdot (E_\a(1) \otimes \hat \gamma_{-\a} + 1
\otimes
       E_\a(1)) $$ on $U_1$. Since $U \hat \otimes U$ is a subalgebra
of $(A \otimes A)^*$, the multiplicativity of $\D$ and the fact that
$U$ is generated by $U_1$ implies that $\D U \in U \hat \otimes U$.
Moreover, this gives us an explicit description of the comultiplication.
\medskip
So the category
${\cal M} (U,H)$ becomes the tensor category; and we have
a monoidal functor ${\rm Rep}(A) \ra {\cal M}(U,H)$.
\medskip
{\bf *.3.4.} {\it Resume.} We choose a subalgebra $U \subset
A^*$ which is attentive to Hopf ideal $I$. Then we complete
$U \otimes U$ in order to define $\D$. This completion is
easily described in terms of the completion $S^* \otimes
S^*$ to $(S \otimes S)^*$. Then we restrict ourselves to
the category  ${\cal M}(U,H)$ which is our choice of dual object
to $A$.
\medskip
{\bf *.3.5}  Later in this paper instead of $U$ we will
consider a smaller  algebra $U^0 \subset U$, which is generated
by $\hat h (h \in H)$ and by elements $E_{\a}(h)$
satisfying the relations:
$$\eqalign {
E_\a(h_1)\hat h_2 & = \a(h_2)E_\a(h_1h_2) \cr
\hat h_2 E_\a(h_1) & = E_\a(h_1h_2) \cr
\D\hat h & = \hat h \otimes \hat h \cr
\D(E_\a(h)) & = E_\a(h) \otimes \hat h \hat \gamma_{-\a}
+ \hat h \otimes E_\a(h)\ .\cr
}$$
The antipode in $U^0$ is defined as follows: $$
i(E_\a(h)) = - \hat h^{-1}E_\a(h)\hat h^{-1} \hat \gamma_{-\a}^{-1}\ .
$$
\bigskip
\noi {\bf *.4. Universal object.}
 \medskip
 {\bf *.4.1.} We denote by $\wtilde U$ a free algebra
 generated by elements $\hat h \ (h
\in H)$
 and by elements $E_\a(h)$ satisfying the relations *.3.5.
\medskip
We define a comultiplication $\D$ on $\wtilde U$ as in *.3.3.
We define the category ${\cal M}(\wtilde U,H)$ of $(\wtilde U,H)$-modules
as in *.3.1. Then we have an epimorphism of Hopf algebras
$\wtilde U \ra U$ which induces a fully faithful monoidal functor
${\cal M}(U,H) \ra {\cal M}(\wtilde U,H)$; and hence, the functor
${\rm Rep}(A) \ra {\cal M}(\wtilde U,H)$.
\medskip
{\bf *.4.2.} {\it Summary.}
We constructed a Hopf algebra $\wtilde U$ using only the $S$-tetramodule $T$.
And for any $S$-tetramodule
$T$ we can construct a Hopf algebra $\wtilde U$.
\bigskip
\noi {\bf *.5. $\bf SL_q(2)$.}
\medskip
{\bf *.5.1.} We apply our approach to the algebra $A_q$ of
functions on the quantum group $SL_q(2)$. The definition of $A
_q$ and $I \subset A_q$ is given
in section 1.
\medskip
The space  $M$ of right $H$-invariants in $I/I^2$ is two-dimensional:
 $M = M_\a \oplus M_{-\a}$, where $\a$ is a character of weight 2. And
$$\g_\a = q^{-1}\quad , \quad \g_{-\a} = q^{-1}\ .$$
\medskip
{\bf *.5.2.} In this case the Hopf algebra $\widetilde U$ is generated by
elements $\hat h, h \in H \approx \C^*$
and by elements $E(h) = E_\a(h)$ and
$F(h) = E_{-\a}(h)$ satisfying the relations:
$$  \leqalignno{
E(h_1) \hat h_2 & = h_2^{-2}E(h_1h_2) &(1)\cr
\hat h_2 E(h_1) & = E(h_1h_2) &\cr
F(h_1) \hat h_2 & = h_2^2 F(h_1h_2) &\cr
\hat h_2 F(h_1) & = F(h_1 h_2) &\cr
}$$
$$\leqalignno {
\D \hat h & = \hat h \otimes \hat h & (2) \cr
\D E(h)   & = E(h) \otimes (\what{q^{-1}} \hat h) + \hat h \otimes E(h) & \cr
\D F(h)   & = F(h) \otimes (\what{q^{-1}} \hat h) + \hat h \otimes F(h)
\ . & \cr
}$$
\medskip
{\bf *.5.3.} Let us see how the algebra $\widetilde U$ constructed from $A_q$
corresponds to the definition of the enveloping algebra of
quantum  group  $SL_q(2)$
(see [D],[J], we use notations of [L]).
\medskip
We denote by $K$ - an element
 in $\widetilde U$ which corresponds to $\hat q \in H,\ K =
\hat q \in S^*$.
$$E = E(q), \ F = F(1) \ .$$
Then
$$K E K^{-1} = q^2 E \qquad KFK^{-1} = q^{-2} F $$
$$ \eqalign {
\D K &   = K \otimes K \cr
\D E &   = K \otimes E + E \otimes 1 \cr
\D F &   = 1 \otimes F + F \otimes K^{-1}\ . \cr
}$$
\medskip
{\bf *.5.4.} We did not get the formula for $[E,F]$ in the quantum group
$SL_q(2)$. We could not have done it, because we took into consideration in
our axiomatic approach only the structure of $gr_0 A_q \oplus gr_1 A_q$. This
structure forgets, for example, that $bc = cb$.
\medskip
In the next sections we will develop our axiomatic approach
to get a relation
for $[E,F]$.
\bigskip
\noi {\bf 5. Universal objects.}
\medskip
{\bf 5.1.} Denote by
${\cal H}(S,T) $ - the category of graded Hopf algebras
$B$, such that $B_0 = S, B_1 = T$; and $B$ supplies $T$ with the
given $S$-tetramodule structure.
\medskip
{\bf 5.2.} {\it Lemma.}
\medskip
1) The category ${\cal H}(S,T)$ has the initial object $B^i$
such that for any object $B$ there exists a canonical morphism
$B^i \ra B$;
\medskip
2) The category ${\cal H}(S,T)$ has the final object $B^f$
such that for any object $B$ there exists a canonical morphism
$B \ra B^f$;
\medskip
3) The category ${\cal H}(S,T)$ has the minimal object $B^m$
such that for any object $B$ there exists a subalgebra
$B^\prime \subset B$ and a canonical epimorphism
$B^\prime  \ra B^m$.
\medskip
{\it Proof.} It is easy to check that the construction
in following paragraphs gives objects in question.
\medskip
{\bf 5.3.}
Given an $S$-tetramodule $T$ we can consider $T$ as an $S$-bimodule
and construct a universal
graded algebra $B^i(S,T)$, such that $B^i_0 = S$,
and $B^i_1 = T$; and $B^i$ supplies $T$ with the
given $S$-bimodule structure. $B^i$ is
a universal object as an algebra. Coalgebra structure is reconstructed
on $B^i$ by $S$-comodule structure on $T$ and multiplicativity.
The antipode is uniquely reconstructed on $B^i$ by the antipode
on $S$ and its properties (see
tensor algebra in [W]).
\medskip
{\bf 5.4.} Now, given an $S$-tetramodule $T$ we would like to
construct a universal object with respect to coalgebra structure
of $T$.
\medskip
Given two $S$-bicomodules $V_1$ and $V_2$
we denote by $V_1 \otimes^S V_2$ a subspace in $V_1 \otimes
V_2$ so that it is a kernel of an operator:
$$ c_r \otimes id - id \otimes c_{\ell}: V_1
\otimes V_2 \ra V_1 \otimes S \otimes V_2$$
(see [Kh], or cotensor product in [Sch]).
\medskip
Given an $S$-bicomodule $V$ denote $B^f_n(S,V)$
the space $(...(V \otimes^S V) \otimes ... \otimes^S V)$
($n$ times). It is easy to see that $B^f_n$ is
an $S$-bicomodule and is isomorphic to $B^f_k
\otimes^S B^f_m$ for $k + m = n; \ k,m \ge 0$.
\medskip
Denote $B^f (S,V) = \build{\Sigma}_{0 \le n}^{} B^f_n$.
The space $B^f$ is supplied with the natural
comultiplication structure. Namely, we define
$\D B^f_n \ra B^f \otimes^S B^f
\subset B^f \otimes B^f$ so that for
any $k,m \ge 0 \ k + m = n$ the composite map
$$\D B^f_n \ra B^f \otimes ^S
B^f \build{\ra}_{}^{pr \otimes pr} B^f_k
\otimes B^f_m$$
would be a canonical isomorphism.
\medskip
{\it Proposition.} Above defined $B^f (S,V)$ is a
universal graded coalgebra with $B^f_0 = S, \ B^f_1 = V$;
and $B^f$ supplies $V$ with the given $S$-bicomodule
structure.
\medskip
{\bf 5.5.} {\it Statement.} Given an $S$-tetramodule $T$ the
universal coalgebra $B^f (S,T)$ is supplied with
canonical Hopf algebra structure.
\medskip
{\it Proof.} Actually, Hopf algebra $B^f$ is universal
in a stronger sense: namely, for any graded Hopf algebra
$C$ and compatible morphisms $C_0 \ra S, \ C_1 \ra T$;
there is a canonical morphism $C \ra B^f(S,T)$.
It is easy to see that $B^f(S,T) \otimes B^f(S,T)$ is
a universal free graded coalgebra $B^f(S\otimes S, S \otimes T +
T \otimes S)$.  We have natural morphisms: $m: S \otimes S
\ra S$ and $c_{\ell} + c_r: S \otimes T + T \otimes S \ra T$.
By universality property they define a multiplication map:
$B^f(S,T) \otimes B^f(S,T) \ra B^f(S,T)$.
\medskip
The antipode could be always defined on $T$ and by
universality property reconstructed on $B^f$.
\medskip
{\bf 5.6.} By definition of $B^i (B^f)$, we have a canonical
morphism $\bar \D$:
$$\bar \D: B^i(S,T) \ra B^f(S,T).$$
We put $B^m$ equal to ${\rm Im} \bar \D$. For any object $B$
we can canonically write $\bar \D$ as the composition map:
$B^i \ra B \ra B^f$. This means that $B$ has a subalgebra
$B^\prime = {\rm Im} B^i $, such that $B^m$ is a
surjective image of $B^\prime$. End of proof of Lemma 5.2.
\medskip
{\bf 5.7.} {\it Comments.} 1)The Hopf algebra $B^f$
is in natural duality with the Hopf algebra
$\widetilde U$ described in *.4. The morphism $\widetilde U
\ra U$ corresponds to a morphism $A \ra \hat B^f$,
where $\hat B^f$ is the natural completion of $B^f$
(namely, $\hat B^f = \prod_i B^f_i$).
\medskip
2) The construction of a minimal object $B^m$ in the category
${\cal H} (S,T)$ is very similar to general construction
of irreducible representations in representation theory.
For example, in case of category ${\cal O}$ one describes
explicitly basic modules $M_\lambda$ and $\delta(M_\lambda)$;
and then describes an irreducible object $L_\lambda$ as the image
of the canonical morphism $i: M_\lambda \ra \delta (M_\lambda)$.
Similar construction appears in Langlands classification of
representations of real reductive groups (and, on more
elementary level, in classification of representations of
symmetric groups).
\bigskip
\noi {\bf 6. Back to $\bf A_q $.}
\medskip
{\bf 6.1.} In this and following sections we would like to describe
the situation discussed above in $SL(2)$-case. Namely,
consider
\medskip
{\it Assumption III.} Our torus $H$ is one-dimensional; and
the space $T^H$ of right
$H$-invariants is the sum of two one-dimensional spaces
of weights $\a$
and $-\a$.
\medskip
{\bf 6.2.}
Consider an object $B$ in the category ${\cal H}(S,T)$.
By definition $B_0 = S, \ B_1 = T$.
We would like to discuss now what we can tell about $B_2$.
\medskip
Consider the algebra $B^i(S, T)$.
The space $B^i_2$
would be
isomorphic to $T \otimes_S T$. The space of right $H$-invariants
in $B^i_2$ is four-dimensional
of weights $(-2\a, 0, 0, 2\a)$. The same is true for
$B^f(S,T)$.
\medskip
Consider the algebra $B = gr A_q$, where $A_q$ is the Hopf
algebra of functions on $SL_q(2)$. Then $B_2 = I^2/I^3$.
Its space of right $H$-invariants is three-dimensional and corresponds to
three one-dimensional representations of $H$ of weights $(-2 \a,\ 0,\ 2 \a)$.
Hence, the canonical morphism $B^i \ra gr A_q$ has a kernel.
This comes from the fact that in $A_q$ (and, hence, in $gr A_q$) we have
the relation $bc = cb$ (see 1.1 and *.5.4); while in $B^i$
this relation is absent.
\medskip
{\bf 6.3.} We see that in $SL(2)$ case $B^m$ is smaller than
$B^i$ and $B^f$ ($\bar \D$ has a kernel). In fact, $B^m$ is much smaller,
since $B^i$ has exponential growth -- ${\rm dim} (B^i_k)^H = 2^k$,
while $gr A_q$ has polynomial growth -- ${\rm dim}
(gr_k A_q)^H = k + 1$. This shows that it is important
to investigate tetramodules for which $\bar \D$ has
a nontrivial kernel.
\medskip
{\bf 6.4.} To perform the calculations
we would like to restrict
ourselves to the study of the second graded component
of $\bar \D (B^i
\ra B^f)$:
$$\bar \D_2
: T \otimes _S T \ra T \otimes^S T.$$
The existence of the
 kernel gives us a hope to get a Hopf algebra similar to
$A_q$.
\medskip
{\bf 6.5.} So we come to a
\medskip {\it Definition.} We will call an $S$-tetramodule $T$ an
$S$-tetramodule of $SL(2)$-type if $S = \C[H],\ (H = \C^*),\ T^H =
M_\a \oplus  M_{-\a}$  and the
operator $\bar \D$ on $(T
\otimes_S T)^H$ has one-dimensional kernel of weight $0$.
\bigskip
\noi {\bf 7. S-tetramodules of SL(2)-type.}
\medskip
{\bf 7.1.} Let $T$ be an $S$-tetramodule of $SL(2)$-type.
Let $s \in S$ be the coordinated function on $H \cong \C^*$.
We have $T^H = M_\a \oplus M_{-\a}$. The weight $\a$ equals $n:
 \a = s^n$. Bimodule structure of $T$ is described in 4.3.
 It is defined by two points of the torus $\gamma_\a$ and
 $\gamma_{-\a}$. In our case they correspond to two numbers
 $s(\gamma_\a)$ and
 $s(\gamma_{-\a})$ in $\C^*$.
Let us choose $t_1 $ and
$t_2$ elements of $M_\a$ and $M_{-\a} $ respectively. We have
$$ \eqalign {
\D t_1   & =  s^n \otimes t_1 + t_1 \otimes 1      \cr
\D t_2   & =  s^{-n} \otimes t_2 + t_2 \otimes 1    \cr
s t_1    & = s(\g_\a) t_1 s        \hfill  \cr
s t_2    & = s(\g_{-\a})t_2  \hfill  \ .
}$$
\medskip
The space of right $H$-invariants in $T \otimes_S T$ of
weight $0$ is spanned by $t_1t_2$ and $t_2t_1$. We apply $\bar \D$ to
these elements:
$$\eqalign{
\bar \D (t_1t_2) &= \a(\g_{- \a}) t_2 s^n \otimes
t_1 + t_1 s^{-n}
\otimes t_2 \cr
\bar \D (t_2 t_1) &= t_2 s^n \otimes t_1 + (-\a) (\g_\a)
t_1 s^{-n}
\otimes t_2 \ . \cr
}$$
\medskip
{\bf 7.2.}  Simple  calculation  immediately  gives   us   the
following
\medskip
{\it Lemma.} $T$ is an $S$-tetramodule of $SL(2)$-type iff
$\a(\g_\a) = \a (\g_{-\a})$, or equivalently,
$s(\g_\a)^n = s(\g_{-\a})^n$.
\medskip
{\bf 7.3.} Let us denote
$s(\g_{-\a})^{-1}$ by $q$. Then $s(\g_\a)=\ve q^{-1}$, where
$\ve$ is an $n$-th root of unity. For generic $q$
the kernel of $\bar \D_2$
is spanned by the vector
$t_1t_2 - q^{-n} t_2t_1$. It is easy to see that
this vector generates a Hopf ideal in
$B^i$. So we can take a quotient of $B^i$ by this ideal.
\medskip
We will denote
the quotient algebra by $B_q(n,\ve)$.
\medskip
{\bf 7.4.} {\it Remark.} Understanding that the  kernel could
be of some importance, let us
 check whether  $\left .\bar \D \right |_{(T
\otimes_S T)^H}$ has
 kernels of weights $2 \a$ or $(-2 \a)$. It is
easy to see that there exists a kernel of weight $2 \a \ ({\rm resp.}
\ -2 \a)$ iff $s(\g_\a)
^n =  -  1 \ ({\rm resp.} \ s(\g_{-\a})^n = -1)$.  Hence we can expect
that the
points $q \ (q^n =
-1)$ are special for $B_q (n,\ve)$.
\medskip
{\bf 7.5.} {\it Example.} $gr A_q \approx B_q (2,1)$. Special points are
$q = \pm i$.
\bigskip
\noi {\bf *.6. Dual picture. Quantum commutator.}
\medskip
{\bf *.6.1.} We want to construct a Hopf algebra  dual  to  $B_q
(n,\ve)$. In *.4 we have described an algebra $\widetilde U$ which is
dual to $B^f(n,\ve)$. The algebra $\widetilde U$ is generated
by elements
$\hat h,
 (h  \in  H)$
and  by elements $E(h) = E_{\a}(h)$ and
$F(h) = E_{-\a}(h)$ satisfying the relations:
$$ \leqalignno{
E(h_1) \hat h_2 & = s(h_2)^{-n} E(h_1h_2) &(1)\cr
\hat h_2 E(h_1) & = E(h_1h_2) &\cr
F(h_1) \hat h_2 & = s(h_2)^n F(h_1h_2) &\cr
\hat h_2 F(h_1) & = F(h_1h_2) &\cr
}$$
$$ \leqalignno{
\D \hat h & = \hat h \otimes \hat h &(2)\cr
\D E(h) & = E(h) \otimes (\hat q^{-1} \hat h) + \hat h \otimes
E(h) &\cr
\D F(h)  & = F(h) \otimes (\hat \ve \hat q^{-1} \hat h) + \hat
h \otimes F(h) &\cr
}$$
\medskip
{\bf *.6.2.} To get a quotient algebra in  $\widetilde  U$  which  is
dual to $B_q(n,\ve)$, we have to add the relation in $\widetilde U^2$
which is orthogonal to the image of $\bar \D_2$.
That is $$F(1)E(1) - q^{-n} E(1) F(1) = 0. $$
This relation is equivalent to the relation
$$ s(h_2)^n E(h_1) F(h_2) - q^n s(h_1)^{-n} F(h_2) E(h_1) = 0 $$
for any $h_1, h_2 \in H$.
\medskip
{\bf *.6.3.} For any Hopf algebra $B$, which is an $A$-tetramodule
we can define a quantum commutator on $B$ with
respect to $A$: $[\ ,\ ]_A$ (we would use this definition when
$B$ is graded Hopf algebra and $A = B_0$).
This definition is due to Woronowicz
(see external algebra in [W]). Namely, we have the standard
adjoint action of $A$ on $B$:
$${\rm Ad}_A a: A \otimes B \ra B \ \ \ \ b \mapsto
\sum_k a_k^1 \cdot b \cdot i(a_k^2),$$ where $\D a = \sum_k a_k^1
\otimes a_k^2$.
\medskip
{\it Definition.} We define $[\ ,\ ]_A$ as the
difference of two operators from $B\otimes B$ to $B$:
$$B \otimes B \build{\ra}_{}^{id \otimes c_\ell}
B \otimes A \otimes B \build {\ra}_{}^{S_{12}} A \otimes B \otimes B
\build {\ra}_{}^{Ad_A \otimes id} B \otimes B \build {\ra}_{}^{m} B$$
$$B \otimes B \build{\ra}_{}^{c_r \otimes id}
B \otimes A \otimes B
\build {\ra}_{}^{id \otimes Ad_A} B \otimes B \build
{\ra}_{}^{S_{12}} B\otimes B \build {\ra}_{}^{m} B.$$
\medskip
{\it Examples.} 1) If $B$ is a tetramodule over the field $k$,
then $[b_1, b_2]_k = b_1b_2 - b_2b_1$.
\medskip
2) If $b_1$ is a right $A$-coinvariant, and $b_2$ is a left
$A$-coinvariant, then $[b_1,b_2]_A = b_1b_2 - b_2 b_1$.
\medskip
{\bf *.6.4.}
We can write the relation in *.6.2 as a quantum commutator
in $\widetilde U$ with respect to $H$ (subalgebra generated by
$\hat h$):
$$\a(h_2) E(h_1)F(h_2)
- q^n (-\a)(h_1) F(h_2) E(h_1) = [E(h_1),F(h_2)]_H$$
\medskip
If
$s(h_1) = q s(h_2)^{-1}$, then the quantum com\-mu\-tator is
proportional to the
usual one:   $[E(qh^{-1}),   F(h)]_H   = h^n
[E(qh^{-1}), F(h)]$,   in    parti\-cular
$[E(q),F(1)]_H    =
\- [E(q),F(1)]$. We have the following important
relation: $$\D[E(h_1),F(h_2)]_H = [E(h_1),F(h_2)]_H
\otimes (\hat \ve \hat q^{-2} \hat h_1 \hat h_2) +
(\hat h_1 \hat h_2) \otimes [E(h_1), F(h_2)]_H,$$
which is equivalent to
$$\D [E(q), F(1)]_H = [E(q), F(1)]_H \otimes \hat \ve
\hat q^{-1} + \hat q \otimes [E(q),F(1)]_H.$$
\medskip
{\bf *.6.5.}  {\it Statement.} The algebra generated by elements
$\hat h, E(h), F(h) \ \ (h \in H)$ and the relations (1), (2) and
$$ [E(q), F(1)]_H = 0 \leqno(3)
$$
is a Hopf algebra.
\medskip
This algebra is "dual" to $B_q(n,\ve)$.
\bigskip
\noi {\bf *.7. Metaplectic quantum groups of SL(2)-type.}
\medskip
{\bf *.7.1.}
     We have  constructed  the   family   of graded  Hopf   algebras
$B_q(n,\ve)$ which are
analogues of  $gr  A_q$.  Our goal is to describe a
family of quantum groups $G = G_q(n,\ve)$ which we call metaplectic
groups of $SL(2)$-type. The algebra $A_q(n,\ve)$ of functions on this group
has the property:
$gr A_q(n,\ve) = B_q (n,\ve)$.
These algebras are the analogues of $A_q$.
We will call $A_q(n,\ve)$ the
algebra of functions on metaplectic quantum group $SL_q(2)(n,\ve)
$ of  $SL(2)$-type.
In this section we construct a Hopf  algebra $U_q(n,\ve)$
which is dual to $A_q(n,\ve)$ and which could be considered
as the  enveloping  algebra  of  metaplectic  quantum   group
$SL_q(2)(n,\ve)$.
     \medskip
     {\bf *.7.2.}
     Using the decomposition $A/I^2 = S \oplus T$ (see 4.5),
     we see that
     $U_(n,\ve)^1 = S^* \oplus T^*$. Thus,
     we have a canonical morphism of Hopf algebras
     $\widetilde U_q (n,\ve) \ra U_q(n,\ve)$.
     Passing to associated graded algebras we have
a morphism $\phi$: $\widetilde U = gr \widetilde U
     \ra gr U = \oplus B^*_i$. As we saw this
     morphism gives us an isomorphism:
     $gr U = \widetilde U/K$, where $K$ is the relation
     $[E(q),F(1)] = 0$.
     This implies that $U = \widetilde U/K^\prime$, where $K^\prime$
     is the relation
     $[E(q),F(1)] +Q = 0$, where $Q \in \widetilde U^1$.
     Since $\phi$ is equivariant with respect to adjoint
     action of $H$ and $K$ is invariant with respect
     to this action, this implies that $Q$ is
     ad-invariant and hence $Q \in \widetilde U_0 = S^*$.
     Since $\phi$ is a morphism of Hopf algebras, then $Q$
     satisfies the relation (see *.6.4):
$$\D Q  =  \hat  q  \otimes Q + Q \otimes \hat \ve \hat q^{-1}.
\leqno (*)$$
\medskip
{\bf *.7.3.}  {\it  Statement.} All solutions of $(*)$ are
proportional to
$$\hat q - \hat \ve \hat q^{-1}.$$
We normalize $Q$ by the condition $Q(s) =  1$,  so  we  will
choose $$Q  =  \frac  {\hat  q - \hat \ve \hat q^{-1}} {q - \ve
q^{-1}}.$$
\medskip
{\bf *.7.4.} {\it Theorem.} The elements $\hat  h,  E(h),  F(h)$
satisfying the relations *.6.1 and the relation
$$[E(q),F(1)]_H = \frac {\hat q - \hat
\ve \hat q^{-1}} {q - \ve q^{-1}}  \leqno (3)$$
generate a Hopf algebra.
\medskip
This is the announced Hopf algebra $U_q(n,\ve)$.
\medskip
{\bf *.7.5.} We denote $E(q)$ by $E$,  $F(1)$ by F.  We  can
rewrite the  definition  of metaplectic quantum groups in the more
compact form:
\medskip
{\it Definition.}  The   Hopf   algebra   $U_q(n,\ve)$   is
generated by multiplicative elements $\hat h  \ (h \in H)$,
and $E$ and $F$ satisfying
the relations:
$$\leqalignno {
\hat h E \hat h^{-1} & = s(h)^n E &(1)\cr
\hat h F \hat h^{-1} & = s(h)^{-n} F \cr
}$$
$$ \leqalignno {
\D \hat h & = \hat h \otimes \hat h &(2)\cr
\D E & = E \otimes 1 + \hat q \otimes E \cr
\D F & = F \otimes \hat \ve \hat q^{-1} + 1 \otimes F \cr
}$$
$$
[E,F] =  \frac {\hat q - \hat \ve \hat q^{-1}} {q - \ve q^{-1}}
\leqno (3)$$
\medskip
{\bf *.7.6.}  {\it Examples.} 1.The subalgebra of $U_q(2,1)$,
generated by $E,F,  K = \hat q$ gives us the usual  definition
of universal  enveloping  algebra  of  quantum group $SL_q(2)$
(see [L]). We think that  our  approach  has  some  advantages.  For
example, we  have  only  one one-dimensional representation of
our algebra -- the trivial one.
\medskip
2.The subalgebra of $U_q(1,1)$ generated by the $E,F,  K =
\hat q$ gives us the usual definition of universal enveloping
algebra of quantum group $PSL_q(2)$.
\medskip
3. The Hopf algebra $U_q(2,-1)$ could be described as the specialization
of $GL_{p,q}(2)$ (see [M2]), when $p = -q$.
\bigskip
\noi {\bf 8. Hopf algebra of regular functions on metaplectic
quantum groups of SL(2)-type.}
\medskip
{\bf 8.1.} {\it Axiomatic approach.} We would like now to describe
the
algebra $A_q(n,\ve)$ of functions on the metaplectic group
$G_q(n,\ve)$. This should be a Hopf algebra
dual to $U_q(n,\ve)$ such that
$gr A_q(n,\ve) = B_q(n,\ve)$.
\medskip
In general we are not sure whether in all cases it is possible
to construct an algebra $A$ which is a Hopf algebra in
a standard sense. The reason is that the only things we really
can describe with our approach are the
quotients $A/I^n$ for all $n$. This means that the algebra
which we can construct is the completion of $A$ with respect
to powers of ideal $I$. In other words, the natural dual object
to $U$ will be an algebra $A$ with a two-sided ideal $I$ complete
with respect to the I-adic topology defined by powers $I^n$ of the
ideal $I$.
\medskip
What about comultiplication. If we want it to be a morphism
$\D: A \to A \otimes A$, then we do not know how to construct it
(in fact there is now reason to expect that the comultiplication
will be defined on the whole completed
algebra $A$). However, working in this situation it is natural to
replace $A \otimes A$ by a completed tensor product $A \hat \otimes
A$. Namely, consider a two-side ideal $I^\prime = I \otimes A +
A \otimes I$ of the algebra $A \otimes A$ and denote by $A \hat
\otimes A$ the completion of $A \otimes A$ in $I^\prime$-adic topology.
\medskip
{\bf 8.2.}
It turns out, that with this definition we can define on $A$ the
natural structure of a completed Hopf algebra. In other words,
we will construct a pair $(A,I)$, where $I$ is a two-sided ideal
of $A$ and $A$ is completed in $I$-adic topology, and a comultiplication
$\D: A \to A \hat \otimes A$, which satisfies all the axioms of Hopf
algebra.
\medskip
Namely, consider the completed algebra $\hat B^f$ corresponding
to our tetramodule $T = T(n,\ve)$. It is easy to see that it is
a completed Hopf algebra in a sense described above, and that it is
in natural duality with the algebra $\widetilde U(n,\ve)$. Now
let us denote by $K^\prime$ the kernel of the natural projection
$\widetilde U (n, \ve) \to U(n,\ve)$ described in *.7.2 and define
$A(n,\ve) \subset \hat B^f$ to be the orthogonal complement
of $K^\prime$,
i.e. $A = \{b \in B| (b,K^\prime) = 0\}$.
\medskip
Since $\widetilde U(n,\ve) \to U(n,\ve)$ is a morphism of Hopf algebras,
$A$ is a completed Hopf algebra dual to $U(n,\ve)$.
\bigskip
\noi {\bf *.8. Representations of $\bf U_q(n,\ve)$.}
\medskip
{\bf *.8.1.} Let $U = U_q(n,\ve)$ be the universal
enveloping algebra of metaplectic group $G_q(n,\ve)$.
We denote by ${\cal M}(G)$
the category ${\cal M}(U,H)$ of $(U,H)$-modules.
We consider the subcategories ${\cal F}(G) \subset
{\cal O}(G) \subset {\cal M}(G)$, where ${\cal F}$
is the category of finite dimensional modules and
${\cal O}$ is the category of ${\cal O}$-modules, i.e.
finitely generated $(U,H)$-modules $M$ for which
weight spaces $M_i$ are $0$ for large $i$.
\medskip
The comultiplication in $U$ allows to define a tensor product
of $(U,H)$-modules and hence defines on each of this
categories the structure of monoidal category.
\medskip
{\bf *.8.2.} Let $l$ be a divisor of $n$. Then the group
$H$ has a unique cyclic subgroup $C$ of order $l$.
It is clear from our definitions that this subgroup
lies in the center of the algebra $U(n,\ve)$. In other words,
$C$ can be considered as the central subgroup
in the metaplectic quantum group $G_q(n,\ve)$.
\medskip
We can consider a quotient quantum group $G/C$. Its universal
enveloping
algebra $U(G/C)$ equals $U/I$,where $I$ is the
ideal generated by $\{ \hat c - 1 | c \in C \} $.
The corresponding algebra of functions $A(G/C)$
can be described as a subalgebra of $C$-invariant
elements in $A(G)$ with respect to the natural left
(equal to right) action of $C$ on $A(G)$.
It is easy to see that $U(G/C)$ is the algebra of
type $U (n/l, \ve^l)$. In particular, we have the natural
monoidal functor ${\cal M}(G/C) \to {\cal M}(G)$, which
transforms ${\cal O}$-modules into ${\cal O}$-modules
and finite dimensional modules into finite dimensional ones.
\medskip
{\it Examples.} 1) Let $l = n$. Then $G/C$ is isomorphic to
$PSL_q(2)$.
\medskip
2) Let $n$ be even, $l = n/2$. If $\ve^l = 1$ then $G/C$ is
isomorphic to $SL_q(2)$.
\medskip
{\bf *.8.3.} Let $l$ be the minimal positive multiple of $n/2$
such that $\ve^l = 1$, that is $l = n$ if $n$ is odd, or $n$ is even
and $\ve^{n/2} = -1$; $l = n/2$ if $n$ is
even and $\ve^{n/2} = 1$.
\medskip
{\it Proposition.} Let $C$ be the central subgroup, corresponding
to $l$. Then for generic $q$ any finite dimensional
representation is trivial on
the subgroup $C$. In other words, the functor: ${\cal F}(G) \to
{\cal F}(G/C)$ is an equivalence of the categories.
\medskip
Thus, the monoidal category of finite dimensional representations
of a metaplectic group $G$ is equivalent to the category of
finite dimensional representations of one of the classical groups
$SL_q(2)$ or $PSL_q(2)$.
\medskip
{\it Proof.} Let $V$ be a finite dimensional $(U,H)$-module. We can
assume that it is irreducible. Standard argument shows that $V$ has
a highest weight vector $v_0$, such that $v_0$ has the weight $N$,
and $E v_0 = 0$. Let us denote $F^i v_0$ by $v_i$. The weight of
$v_i$ equals $N - ni$. The commutation relation for $E$ and $F$
shows that
$$E v_i = [i]_{q^{n/2}}(q^{N - n(i-1)/2} -
\ve^N q^{-N + n(i-1)/2})/(q - \ve q^{-1}),$$
where $[i]_a = (q^a - q^{-a})/(q - q^{-1})$.
\medskip
There exists $k \ne 0$ such that $v_k \ne 0$ and $F v_k = 0$.
Then the vectors $v_0, \dots , v_k$ generate $k+1$-dimensional
irreducible representation of $G$. Thus for generic $q$ we have
that $2N = nk$ and $\ve^N = 1$. Hence $N \in l\Z$.
\medskip
{\bf *.8.4.} Let $C \subset H$ be a central subgroup in $G$.
For any character $\xi$ of $C$ we denote by ${\cal M}_\xi (G)$
the subcategory of modules on which $C$ acts via character $\xi$.
Clearly ${\cal M}(G) = \oplus_\xi \ {\cal M}_\xi (G)$ and
${\cal M}_\xi \otimes {\cal M}_\eta \subset {\cal M}_{\xi \eta}$.
Similarly for ${\cal O}(G)$ and ${\cal F}(G)$.
\medskip
{\bf *.8.5.} Fix a subgroup $C$ as in *.8.3. Then the category
${\cal F}(G)$ is one of two classical categories. For any character
$\xi$ we have compatible right and left actions of this classical
category ${\cal F}(G) = {\cal F}(G/C)$ on the category
${\cal M}_\xi (G)$ and, in particular, on the category
${\cal O}_\xi (G)$.
\bigskip
\noi {\bf References}
\medskip
[A] E.Abe, \it Hopf algebras, \rm Cambridge University Press,
Cambridge - New York, 1977.
\medskip
[D] V.G.Drinfeld, \it Quantum groups, \rm Proceedings of the ICM,
p. 798-820, Rhode Island, AMS, 1987.
\medskip
[G-S] M.Gerstenhaber and S.D.Schack, \rm Algebras, bialgebras, quantum
groups, and algebraic deformations, \rm Contemp. Math. \bf 134
\rm (1992), 51-92.
\medskip
[J] M.Jimbo, \it A q-analog of $U(gl(N+1))$ Hecke algebra, and the
Yang-Baxter equation, \rm Lett. Math. Phys. \bf 11 \rm (1986),
247-252.
\medskip
[Kh] T.Khovanova, \it Tetramodules over the Hopf
algebra of regular functions on a torus, \rm IMRN \bf
\rm (1994)     / hep-th 9404043
\medskip
[L-S] V.Lyubashenko and A.Sudbery, \it Quantum supergroups of $GL(n|m)$
type: Differential forms, Koszul complexes and Berezinians,
\rm / hep-th 9311095
\medskip
[L] G.Lusztig,  \it On quantum groups,  \rm J. of Alg.  \bf 131
\rm (1990), 466-475.
\medskip
[M] Yu.I.Manin,  \it  Quantum   groups   and   non-commutative
geometry, \rm CRM, Universit\'e de Montr\'eal, 1988.
\medskip
[M2] Yu.I.Manin, \it Multiparametric Quantum Deformation
of the General Linear Supergroup, \rm Commun. Math. Phys,.
\bf 123 \rm (1989), 163-175.
\medskip
[R-T-F] N.Yu.Reshetikhin, L.A.Takhtadzhyan and L.D.Faddeev,
\it Quantization of Lie groups and Lie algebras, \rm
Leningrad Math. J. \bf 1 \rm (1990), 193-225.
\medskip
[Sch] P.Schauenburg, \it Hopf modules and Yetter-Drinfel'd
modules, \rm to appear in J. Alg.
\medskip
[S-S] S.Shnider and S.Sternberg, \it Quantum groups,
\rm International Press, Boston - Hong Kong, 1993.
\medskip
[S] M.E.Sweedler, \it Hopf algebras, \rm Benjamin, New York, 1969.
\medskip
[W] S.L.Woronowicz, \it Differential Calculus on Compact Matrix
Pseudogroups (Quantum groups), \rm Commun. Math. Phys.,
\bf 122 \rm (1989), 125-170.
\medskip

                           \end